\documentclass[fleqn,11pt]{article}

\usepackage{amsmath,amssymb,amsfonts}
\usepackage{graphicx}
\usepackage{color}

\usepackage{amsfonts,amssymb,cite}
\usepackage{graphicx}

\def\onecol{\onecolumn \mathindent 2em}

\def\noi{\noindent}

\newcommand{\Title}[1]{\noi {{\Large\bf #1}}\\[1ex]}

\def\Aunames#1{\noi{\bf #1}}
\def\au#1{${}^{#1}$}
\def\Addresses#1{\medskip\noi \protect
	\begin{description}\itemsep -3pt {\it #1} \end{description}}
\def\adr#1#2{\item[${}^{#1}$]{\it #2}}

\newcommand{\Abstract}[1]{\vskip 2mm \begin{center}
        \parbox{16.4cm}{\small\noi #1} \end{center}\medskip}

\def\email#1#2{\footnotetext[#1]{e-mail: #2}\addtocounter{footnote}{1}}

\def\nqq{\hspace*{-2em}}

\def\cm{\hspace*{1cm}}
\def\inch{\hspace*{1in}}
\def\wide{\mbox{$\dst\vphantom{\int}$}}

\def\ten#1{\mbox{$\times 10^{#1}$}}
\def\deg{\mbox{${}^\circ$}}                     

\def\Jl#1#2{#1 {\bf #2},\ }

\def\ApJ#1 {\Jl{Astroph. J.}{#1}}
\def\CQG#1 {\Jl{Class. Quantum Grav.}{#1}}
\def\DAN#1 {\Jl{Dokl. AN SSSR}{#1}}
\def\GC#1 {\Jl{Grav. Cosmol.}{#1}}
\def\GRG#1 {\Jl{Gen. Rel. Grav.}{#1}}
\def\JETF#1 {\Jl{Zh. Eksp. Teor. Fiz.}{#1}}
\def\JETP#1 {\Jl{Sov. Phys. JETP}{#1}}
\def\JHEP#1 {\Jl{JHEP}{#1}}
\def\JMP#1 {\Jl{J. Math. Phys.}{#1}}
\def\NPB#1 {\Jl{Nucl. Phys. B}{#1}}
\def\NP#1 {\Jl{Nucl. Phys.}{#1}}
\def\PLA#1 {\Jl{Phys. Lett. A}{#1}}
\def\PLB#1 {\Jl{Phys. Lett. B}{#1}}
\def\PRD#1 {\Jl{Phys. Rev. D}{#1}}
\def\PRL#1 {\Jl{Phys. Rev. Lett.}{#1}}

\def\lal{&&\nqq {}}
\def\eq{Eq.\,}
\def\eqs{Eqs.\,}
\def\beq{\begin{equation}}
\def\eeq{\end{equation}}
\def\bear{\begin{eqnarray}}
\def\bearr{\begin{eqnarray} \lal}
\def\ear{\end{eqnarray}}
\def\earn{\nonumber \end{eqnarray}}

\def\nnn{\nonumber\\ \lal }
\def\nnnv{\nonumber\\[5pt] \lal }
\def\yy{\\[5pt] {}}
\def\yyy{\\[5pt] \lal }

\def\dst{\displaystyle}

\def\e{{\,\rm e}}

\def\diag{\mathop{\rm diag}\nolimits}

\def\const{{\rm const}}
\def\eps{\varepsilon}
\def\ep{\epsilon}

\def\then{\ \Rightarrow\ }

\newcommand{\aver}[1]{\langle \, #1 \, \rangle \mathstrut}

\def\eqn#1{\eq\eqref{#1}}
\def\rf{\eqref}
\def\mn{_{\mu\nu}}
\def\MN{^{\mu\nu}}
\def\mN{_\mu^\nu}
\def\nM{_\nu^\mu}

\def\R{{\mathbb R}}
\def\S{{\mathbb S}}
\def\cK{{\cal K}}

\def\ob{{\bar b}}
\def\oc{{\bar c}}

\def\kappa{\varkappa}

\def\KS{Kantowski-Sachs}
\def\Swz{Schwarz\-schild}
\def\sph{spherically symmetric}
\def\ssph{static, spherically symmetric}

\def\bh{black hole}
\def\bhs{black holes}

\begin{document}
\thispagestyle{empty}
\onecol

\Title{Matter accretion versus semiclassical bounce\yy
         in Schwarzschild interior }

\Aunames{K. A. Bronnikov,\au{a,b,c,1} S. V. Bolokhov,\au{b;2} and M. V. Skvortsova\au{b;3}}

\Addresses
	{\small
\adr a {Center of Gravitation and Fundamental Metrology, VNIIMS, 
		Ozyornaya ul. 46, Moscow 119361, Russia}
\adr b {Institute of Gravitation and Cosmology, Peoples' Friendship University of Russia (RUDN University),\\
		ul. Miklukho-Maklaya 6, Moscow 117198, Russia}
\adr c{National Research Nuclear University ``MEPhI'', 
		Kashirskoe sh. 31, Moscow 115409, Russia}
        }

\Abstract
  {We discuss the properties of the previously constructed model of a Schwarzschild black hole
  interior where the singularity is replaced by a regular bounce, ultimately leading to a white hole.
  The model is semiclassical in nature and uses as a source of gravity the effective stress-energy 
  tensor (SET) corresponding to vacuum polarization of quantum fields, and a minimum spherical radius 
  is a few orders of magnitude larger than the Planck length, so that the effects of quantum gravity
  should be still negligible. We estimate the other quantum contributions to the effective SET, caused 
  by a nontrivial topology of spatial sections and particle production from vacuum due to a nonstationary
  gravitational field and show that these contributions are negligibly small as compared to the SET
  due to vacuum polarization. The same is shown for such classical phenomena as accretion of 
  different kinds of matter to the black hole and its further motion to the would-be singularity.
  Thus, in a clear sense, our model of a semiclassical bounce instead of a Schwarzschild singularity
  is stable under both quantum and classical perturbations.  
  }



\email 1 {boloh@rambler.ru}
\email 2 {kb20@yandex.ru}
\email 3 {milenas577@mail.ru}

\section{Introduction}

  The existence of singularities in various solutions of general relativity (GR) 
  as well as many alternative classical theories of gravity, describing
  \bhs\ or the early Universe, is an undesirable but apparently inevitable feature. 
  On the other hand, one can hardly believe that the curvature invariants or 
  the densities and temperatures of matter that appear in such
  singularities can really reach infinite values. There is therefore a more or less common
  hope that a future theory of gravity valid at very large curvatures, high energies, 
  small length and time scales will be free of singularities, and that such a theory 
  should take into account quantum phenomena.  
  
  The existing numerous attempts to avoid singularities can be 
  basically classified as follows:
\begin{description}   
\item[(a)] 
   In GR, invoking ``exotic'' sources of gravity violating the standard energy conditions, 
   for example, phantom scalar fields; in classical extensions of GR, 
   using quantities of geometric origin (torsion, nonmetricity, extra dimensions)
   whose effective stress-energy tensors (SETs) can have similar ``exotic'' properties 
   \cite{vis-book,BR-book,lobo-rev,kb-fab-02,kb-NED-01,dym03,guen13,pha1,bu-06,we-12,wh-16};  
   it has also been argued that the effects of rotation in GR can also play the role of 
   exotic matter, see, e.g., \cite{rot1, rot2, rot3}.
\item[(b)] 
   In semiclassical gravity, where the geometry is treated classically and obeys
   the equations of GR or an alternative classical theory, using averages of quantum 
   fields of matter as sources of gravity with possible ``exotic'' properties
   \cite{piran-94,SCG1,garr-07,hiscock-97,corda-11,bardeen-14}.
\item[(c)]
   Diverse models of quantum gravity are also often translated into the language of
   classical geometry and lead to nonsingular space-times describing both regular \bh\
   interiors and early stages of the cosmological evolution
  \cite{QG1, QG2, QG3, kelly20, achour20, kunst08, kunst20, dadh15, ash20a, ash20b, bambi13}
\end{description}
  (see also references therein for each item).

  One can notice that the singularity problems in \bh\ physics and Big Bang cosmology 
  are quite similar. For example, the \Swz\ singularity is located in a nonstationary 
  ``T-region'', where the metric can be written as that of a homogeneous anisotropic
  cosmology, a special case of Kantowski-Sachs models. It is therefore natural that 
  the same tools are used in attempts to attack these problems. 
    
  Classical nonsingular models in cosmology, \bh\ and wormhole physics are quite 
  popular, but the ``exotic'' components that are necessarily present in those models 
  require certain conjectures so far not confirmed by observations or experiments,
  and their consideration is often justified as a kind of phenomenological description of 
  underlying quantum effects.
 
  Many models of quantum gravity, in their representations in the language of 
  classical geometry, lead to nonsingular cosmologies and \bh\ models,
  but most frequently such models reach the values of curvatures and densities close to  
  the Planck scale. However, more surprising is a considerable diversity of their predictions,
  depending on various leading ideas employed in such models. 
  
  Thus, a number of scenarios in the framework of Loop Quantum Gravity (LQG) predict 
  a bounce close to the Planck scale and a transition from a \bh\ to a white hole
  \cite{QG2, QG3, kelly20, achour20}. In particular, in \cite{kelly20, achour20} the authors
  consider quantum corrections to the Oppenheimer-Snyder collapse scenario.
  
  Unlike that, application of the so-called polymerization concept to the interior of a \Swz\
   \bh\ \cite{kunst08, kunst20}, also removing the singularity, leads to a model with a single
  horizon and a \KS\ cosmology with an asymptotically constant spherical radius at late times. 
  (This geometry is partly similar to the classical black universes with a phantom scalar
  \cite{pha1,bu-06,we-12}, but in the latter the late-time \KS\ cosmology tends to de Sitter 
  isotropic expansion.) 
  
  Some of the scenarios (see, e.g., \cite{ash20a}) even lead to a quantum-corrected 
  effective metric with an unconventional asymptotic behavior, although it is claimed that 
  the quantum correction to the \bh\ temperature is quite negligible for sufficiently large
  black holes, and that the metric is asymptotically flat in a precise sense.

  A consideration of homogeneous gravitational collapse of dust and radiation with LQG 
  effects has led in \cite{bambi13} to avoidance of both a final singularity and an event 
  horizon, so that the outcome is a dense compact object instead of a \bh.

  Let us also mention a study of black hole evaporation process by Ashtekar
  \cite{ash20b} using as guidelines (i) LQG, (ii) simplified models with concrete 
  results, and (iii) semiclassical effects. The author discusses various issues concerning
  the information loss problem and the final fate of evaporating \bhs; one of his conclusions 
  is that LQG effects do not appreciably change the semiclassical picture outside 
  macroscopic black holes.
  
  A comprehensive review of quantum gravity effects in gravitational collapse and \bhs\ 
  has been provided by Malafarina \cite{QG1} in 2017, and we here only mention a few 
  results of interest and some papers that appeared later than this review. But
  even this short list shows how diverse can be the results and conclusions depending 
  on the particular approach. All that may be a manifestation of a so far uncertain 
  status of quantum gravity. 
    
  Since matter can manifest its quantum properties at the atomic or macroscopic 
  scales (as exemplified by lasers or the Casimir effect), one may hope that singularities
  in cosmology or \bhs\ may be prevented at length scales much larger than the Planck one.
  This would look more attractive both from the observational viewpoint and also theoretically
  since the corresponding results, at least today, look more confident than those obtained
  with quantum gravity.
 
  The \bh\ studies in the framework of semiclassical gravity, such as
  \cite{piran-94,hiscock-97,corda-11,bardeen-14} 
  and many others, mostly focus on the consequences of the Hawking \bh\ evaporation 
  and the related information paradox. Their conclusions seem promising from the viewpoint
  of singularity avoidance. Thus, in \cite{piran-94} it is concluded that the \bh\ evaporation 
  ultimately leads to emergence of an inner macroscopic region that hides the lost information 
  and is separated from the external world. According to \cite{bardeen-14}, the evaporation
  process even prevents the emergence of an event horizon. Thus, after formation of 
  a large \sph\ \bh\ by gravitational collapse, the classical $r = 0$ singularity is replaced by 
  an initially small regular core, whose radius grows with time due to increasing entanglement 
  between Hawking radiation quanta outside and inside the \bh, and by the Page time 
  (when half the \bh\ mass has evaporated), all quantum information stored in the interior
  is free to escape to the outer space.
 
  However, there remains a question of what is happening inside a large \bh\ when it has just 
  formed, and the evaporation process is too slow to immediately launch the above processes. 
  Indeed, an approximate expression for the full evaporation time is $t_{\rm evap} \propto M^3$,
  where $M$ is the initial \bh\ mass; it then follows that the Page time is $\frac 78 t_{\rm evap}$,
  and if $M$ is the solar mass, we have $t_{\rm evap}\approx 2.1\times 10^{67}$ years.
  In other words, any astrophysical \bh\ (except for very light primordial ones) is at this
  initial stage of its evaporation. Even more than that: under realistic conditions, its mass 
  much faster grows due to accretion than decreases by evaporation. 
  
  In our study we try to answer the following question: what is the internal geometry of such 
  a large and ``young'' \bh\ if its Hawking evaporation can be neglected, but the impact of 
  quantum fields that are present in a vacuum form is taken into account? In other words: if a body 
  (a particle, a planet, a spacecraft) falls into such a \bh, what is the geometry it will meet there?
  
  More specifically, we are considering the neighborhood of a would-be \Swz\ singularity ($r=0$)
  in the framework of semiclassical gravity and explore a possible emergence of a bounce 
  instead of the singularity. We can recall that in any space-time region there always exist
  quantum oscillations of all physical fields. We do not assume any particular composition of 
  these fields, considering only their vacuum polarization effects. In such a simplified statement 
  of the problem, we have shown \cite{we18} that there is a wide choice of the free parameters
  of the model that provide a possible implementation of such a scenario.
  The SET used to describe the vacuum polarization of quantum fields is taken in the form of 
  of a linear combination of the tensors $^{(1)}H\mN$ and  $^{(2)}H\mN$ obtained by variation 
  of the curvature-quadratic invariants $R^2$ and $R\mn R\MN$ in the effective action in 
  agreement with the renormalization methodology of quantum field theory in curved space-times
  \cite{birrell, gmm}.  In this scenario, in the internal \KS\ metric, the spherical radius $r$ 
  evolves to a regular minimum instead of zero, while its longitudinal scale has a regular maximum 
  instead of infinity. The free parameters of the model can be chosen so that the curvature 
  scale does not reach the Planck scale but remains a few orders smaller (for example, on the 
  GUT scale), sufficiently far from the necessity to include quantum gravity effects. The whole
  scenario is assumed to be time-symmetric with respect to the bouncing instant, therefore, as 
  in many other papers, we are describing a smooth transition from black to white hole.
  
  The nonlocal part of the effective SET of quantum fields in the \Swz\ interior, depending on 
  the whole history and mainly represented by particle production from vacuum, was estimated 
  in \cite{we19}, and it was shown that its contribution in the vicinity of a bounce is many
  orders of magnitude smaller than that of $^{(1)}H\mN$ and  $^{(2)}H\mN$. 
  
  In the present paper, after a brief representation of the results of \cite{we18, we19},
  we try to find out whether or not there are classical phenomena that could potentially
  destroy the bounce, namely, accretion of different kinds of matter which is always 
  present near astrophysical \bhs\ and whose density increases as it further moves inside
  the horizon towards the would-be singularity. It turns out that this accretion is also unable
  to affect the bounce due to its negligibly small contribution to the total SET. 
  
  The paper is structured as follows. Section 2 summarizes the problem statement and 
  the assumptions made. In Section 3 we describe the bouncing solution to the field equations,
  in Section 4 we estimate the nonlocal contribution to the effective SET, Section 5 is devoted 
  to calculations of the \sph\ accretion of the CMB radiation and massive matter to a \Swz\
  \bh, and Section 6 is a brief discussion.

\section{Field equations and assumptions}
\subsection{Near-bounce geometry}

   Considering a generic \ssph\ \bh\ in its interior region (beyond the horizon), also called a T-region,
   we can write its metric in the general \KS\ form
\beq       \label{ds0}
	    	 ds^2= d\tau^2 -\e^{2\gamma(\tau)}dx^2-\e^{2\beta(\tau)}d\Omega^2,
\eeq
   where $\tau$ is the natural time coordinate in the corresponding reference frame, and $x$ is 
   a spatial coordinate that ``inherits'' the time coordinate of the static region after crossing the horizon;
   $d\Omega^2$ is, as usual, the metric on a unit sphere $\S^2$. It is a homogeneous anisotropic 
   cosmological model with the topology $\R \times \S^2$ of its spatial sections. 
   
   Assuming that quantum effects can appreciably change the space-time geometry only 
   if the latter is very strongly curved, while at smaller curvatures, even in a T-region  ($r < 2m$), we can  
   use with sufficient accuracy the Schwarzschild solution, which then takes the form  
\beq    				 \label{Swz} 
	  ds^2 = \Big(\frac {2m}{T}-1\Big)^{-1} dT^2  - \Big(\frac {2m}{T}-1\Big)dx^2 
	  	-T^2 d\Omega^2,
\eeq   
  where $m = GM$, $G$ being Newton's constant of gravity, $M$ the \bh\ mass, and 
  we use the units $\hbar=c=1$). Compared to the conventional expression, we have changed 
  the notation, $r \to T$, to emphasize that in the T-region the coordinate $r$ is temporal. 
  Furthermore, at $T \ll 2m$, passing on to the \KS\ cosmological time by putting 
  $\sqrt{T/(2m)} dT = d\tau$, we obtain an asymptotic form of the metric in the notations of \rf{ds0}:
\bearr                     \label {ds1}
	 ds^2 = d\tau^2 - \Big(\frac 43 m\Big)^{2/3} \tau^{-2/3} dx^2 
	 					        - \Big(\frac 92 m\Big)^{2/3} \tau^{4/3}  d\Omega^2,   
\ear      
  which is valid at $\tau/m \ll 1$. It is the \Swz\ metric at approach to the singularity $\tau \to 0$, 
  at which the scale along the $x$ axis is infinitely stretched while the spheres $x=\const$ are 
  shrinking to zero.      
   
  In this study, our basic assumption will be that quantum field effects do not allow the space time
  to approach too close to the singularity $r\equiv \e^\beta = 0$ (or $\tau =0$ in \rf{ds1}) but, 
  instead, stop the contraction of $r$ at $\tau =0 $ at some regular minimum value $r=r_0 > 0$, 
  while the scale factor $\e^\gamma$ along the $x$ axis simultaneously turns to a regular 
  maximum. Then, at small $\tau$, in agreement with \rf{Swz} and \rf{ds1}, the metric takes the form 
\bearr 				\label{ds2}
         ds^2\Bigr|_{\rm bounce} \simeq d\tau^2 - \frac{2m}{r_0}(1- \oc^2\tau^2)dx^2   
         													- r_0^2(1+ \ob^2\tau^2) d\Omega^2
\ear
   where $r_0, \ob, \oc$ are positive constants with appropriate dimensions.

   In addition to these assumptions, let us also suppose that the time evolution of the metric is 
   symmetric with respect to the bouncing instant $\tau=0$.  Then, in the notations of \rf{ds0}, we can
   present the functions $\beta(\tau)$ and $\gamma(\tau)$ as Taylor expansions with only
   even powers of $\tau$, 
\bearr												 \label{beta-gamma}
	\beta(\tau)=\beta_0+\frac{1}{2}\beta_2\tau^2
						+\frac{1}{24}\beta_4\tau^4+\frac{1}{720}\beta_6\tau^6+ \ldots,
\nnn
	\gamma(\tau)=\gamma_0+\frac{1}{2}\gamma_2\tau^2
		+\frac{1}{24}\gamma_4\tau^4+\frac{1}{720}\gamma_6\tau^6+...,
\ear
  where $\beta_i, \gamma_i\; (i=0,2,4,6,...)$ are constants. Then, according to \rf{ds2},
\bearr                                      \label{abc}  
	   	r_0=\e^{\beta_0}, \qquad   	2m/r_0  = \e^{2\gamma_0}, \qquad    		
			    		2 \ob^2 = \beta_2/\beta_0,	\qquad 2 \oc^2 = - \gamma_2/\gamma_0.
\ear

  To explain the behavior  \rf{ds2} of the metric, we invoke the semiclassical approach, writing the Einstein 
  equations as
\beq \label{EE}
		G\mN=-\varkappa \aver {T\mN}, \qquad \kappa = 8\pi G,
\eeq
   where the r.h.s. represents a renormalized  stress-energy tensor (SET) $\aver{ T\mN}$ of 
   quantum fields, containing, in general, both local and nonlocal contributions.

   In the general metric \rf{ds0}, the Einstein tensor $G\mN$ has the following nonzero components:
\bearr          \label{Gmn}
	G^0_0 =  - \dot\beta(\dot\beta+2\dot\gamma) - \e^{-2\beta},
\nnnv
	G^1_1 = - 2\ddot\beta - 3\dot\beta^2 - \e^{-2\beta},
\nnnv
	G^2_2  = G^3_3 = - \ddot\gamma - \ddot\beta - \dot\gamma^2 - \dot\beta^2 - \dot\beta\dot\gamma.
\ear
     Substituting the Taylor expansions \eqref{beta-gamma}, we can explicitly present these components 
     up to $O(\tau^2)$ as follows:
\bearr
	-G^0_0 = \frac{1}{r_0^2}\Big(1-\frac{\beta_2}	{2\beta_0}\tau^2\Big) + \beta_2(\beta_2 + 2\gamma_2)\tau^2,
\nnn
	-G^1_1 = \frac{1}{r_0^2}\Big(1-\frac{\beta_2}{2\beta_0}\tau^2\Big)+2\beta_2+\beta_4\tau^2+3\beta_2^2\tau^2,
\nnn
	-G^2_2 = \beta_2 + \gamma_2 + \frac{1}{2} (\beta_4 + \gamma_4)\tau^2
			+ (\beta_2^2 + \gamma_2^2 + \beta_2\gamma_2)\tau^2.
\ear
 
\subsection{The stress-energy tensor}

  In agreement with the vast literature on quantum field theory in curved space-times, including the books 
  \cite{gmm, birrell}, the renormalized vacuum SET $T\nM$ of quantum fields may be presented as a
  linear combination of two tensors of geometric origin ${}^{(i)}H\mN$ ($i=1,2$) (which can be obtained by 
  variation of actions containing $R^2$ and $R\mn R\MN$, i.e., the Ricci scalar and tensor squared), 
  with some phenomenological constants $N_1, N_2$, and two other contributions,
   ${}^{(c)}\!H\nM$ and $P\nM$:
\beq 					\label{SET}
			\aver{T\nM} = N_1 {}^{(1)}\!H\nM + N_2 {}^{(2)}\!H\nM + {}^{(c)}\!H\nM + P\nM,
\eeq
   where
\bearr                                           \label{HH}
		 {}^{(1)}\!H\nM \equiv 2R R\nM-\frac{1}{2}\delta\nM R^2
								 	+2\delta\nM \Box R-2\nabla_\nu\nabla^\mu R,
\nnn
	{}^{(2)}\!H\nM \equiv
	-2\nabla_\alpha\nabla_\nu R^{\alpha\mu}+\Box R\nM+\frac{1}{2}\delta\nM\Box R
				 +2R^{\mu\alpha} R_{\alpha\nu}-\frac{1}{2}\delta\nM R^{\alpha\beta}R_{\alpha\beta},
\ear
   and $\Box = g^{\mu\nu}\nabla_\mu \nabla_\nu$. The tensor ${}^{(c)}\!H\nM$ is of local nature
   and depends on the space-time topology and/or on boundary conditions (e.g., the Casimir effect 
   \cite{Milton, Elizalde}), while $P\nM$ is nonlocal, it depends on the particular quantum states
   of the constituent fields and, in particular, describes particle production in a nonstationary metric. 
   Its nonlocal nature means that it is not a function of a space-time point but depends, in general, 
   on the whole history. Its calculation is rather a complex task and requires additional assumptions
   on quantum states of different fields. We will temporarily assume that the contribution of $P\nM$ 
   is small as compared to the other terms in \rf{SET} (at least under a suitable choice of quantum state)
   and try to justify this assumption in Section 4. 

  The components of the tensors ${}^{(i)}\!H^\mu_\nu$ (which turn out to be diagonal) can be  
   easily calculated from the ansatz \rf{ds0} with the Taylor expansions \rf{beta-gamma}. At 
   the very instant $\tau =0$ (at bounce) they are
\bearr      \label{Hmn}
            	{}^{(1)}\!H^0_0 = -\frac{2}{r_0^4}+8\beta_2^2+8\beta_2\gamma_2+2\gamma_2^2,
\nnn
          	{}^{(1)}\!H^1_1 = -\frac{2}{r_0^4} \!
         	-32\beta_2^2 - 16\beta_2\gamma_2 - 6\gamma_2^2\! - 8\beta_4 - 4\gamma_4,
\nnn
           	{}^{(1)}\!H^2_2 = 
         	  \frac{2}{r_0^4}+\frac{12\beta_2}{a^2}-24\beta_2^2 - 20\beta_2\gamma_2-10\gamma_2^2
				         	 - 8\beta_4 - 4\gamma_4,
\earn
\bearr
		   	{}^{(2)}\!H^0_0 = -\frac{1}{r_0^4}+3\beta_2^2+2\beta_2\gamma_2+\gamma_2^2,
\nnn
		{}^{(2)}\!H^1_1 = 
		-\frac{1}{r_0^4}-9\beta_2^2 - 6\beta_2\gamma_2-3\gamma_2^2 - 2\beta_4 - 2\gamma_4,
\nnn
	      	{}^{(2)}\!H^2_2 = 
			    \frac{1}{r_0^4} + \frac{4\beta_2}{a^2}-9\beta_2^2 - 6\beta_2\gamma_2 -3\gamma_2^2
				    		 - 3\beta_4 - \gamma_4.
\ear
   As will be clear further on, their higher orders in $\tau$ will be unnecessary in our calculations.
   
   What is known about the numerical coefficients $N_1$ and $N_2$ in \rf{SET}?
   According to \cite{gmm, birrell}, their values should be found from experiments or observations. 
   The orders of magnitude of these coefficients may be roughly estimated by recalling that they 
   appear in higher-derivative theories of gravity where the action has the form 
\beq   
    	S\sim \int d^4x\sqrt{-g}(R/(2\kappa) + N_1 R^2 + N_2 R^2_{\mu\nu}+...)
\eeq    	
   the tensors ${}^{(1,2)}H\mn$ resulting from variation of the corresponding terms.
   The upper bounds on these parameters are $N_{1,2}   \lesssim 10^{60}$ (see, e.g., \cite{Giacchini}), 
   as follows from observations performed at very small curvatures, at which any possible effects of 
   terms quadratic in the curvature are extremely weak. However, the factors $N_{1,2}$ may be 
   estimated in another way if such theories of gravity are used to describe the early (inflationary) 
   Universe with much larger curvatures, for example, $N_1 \sim 10^{10}$ \cite{Starob0, Starob,Odin}. 
   For our purposes, we will keep in mind this order of magnitude.
   
   Concerning the Casimir contribution, there are arguments indicating that it must be much smaller than
   the contribution of ${}^{(i)}H\mN$. If we consider, for instance, the static counterpart of the metric 
   \rf{ds0} with $\e^\beta = r =r_0$, something treatable as a description of an infinitely long wormhole throat,
   we can use the result obtained in \cite{butcher} for a conformally coupled massless scalar field,
   which reads for this geometry
\bearr     \label{Cas}
	   {}^{(c)}\!H\nM =  \frac{1}{2880\pi^2 r_0^4}\Big[ 2\diag(-1,-1,1,1) \ln(r_0/a_0) + \diag(0, 0, -1, -1)\Big], 	   
\ear
  where $a_0$ is some fixed length to be determined by experiment. Note that the quantity \rf{Cas}
  is obtained for a single massless scalar, and the total Casimir contribution must take into account 
  all existing fields with different spins and masses, hence this contribution may be two or three orders
  of magnitude larger than \rf{Cas}.
  
  On the other hand, for the same space-time geometry, 
\beq                                           
             {}^{(1)}\!H\mN = 2\,{}^{(2)}\!H\mN = \frac{2}{r_0^4} \diag (-1, -1, 1, 1).
\eeq    
  Therefore, if $N_1$ and/or $N_2$ are at least of the order of unity (as we will consider in what follows),
  the tensors ${}^{(i)}H\mn$ contribute much stronger to $\aver{T\mN}$ in the Einstein equations \rf{EE} 
  than  ${}^{(c)}H\mn$, unless the uncertain length $a_0$ in \rf{Cas} is unreasonably high, or the total 
  number of fields is so large as to overcome the denominator which is $\sim 10^4$.
   
  In our further consideration we will assume that ${}^{(c)}\!H\nM$ can also be neglected 
  in our geometry \rf{ds2} and take into account only the contributions ${}^{(i)}H\mN$.
  
\section{The semiclassical bounce}

   In this section we consider the Einstein equations \eqref{EE} with the SET \rf{SET}, 
   taking into account only the first two terms. Our task will be to find out whether or not 
   there are solutions consistent with the bouncing metric \rf{ds2}, and if it is the case,
   what are the requirements to the free parameters of the model that would justify 
   the semiclassical nature of the equations. In the subsequent sections we will analyze
   the influence of other effects that could in principle destroy the model thus constructed: 
   the nonlocal contribution to the SET \rf{SET} and the possible influence of matter surrounding the \bh\
   and falling to its interior region.      
  
  For our purpose, we will express $G\nM$ and ${}^{(i)}H\nM$ in terms of the Taylor series
  coefficients in \eqref{beta-gamma} and equate the coefficients at equal powers of $\tau$ on 
  different sides of the resulting equations. Let us introduce, for convenience, the 
  following dimensionless parameters
\bearr                       \label{ABC}
   	A = \kappa r_0^{-2}, \qquad 
   	B_2 = \kappa\beta_2, \qquad 
   	C_2 = \kappa\gamma_2, \qquad  	
   	B_4=  \kappa^2\beta_4,\qquad 
   	C_4 = \kappa^2\gamma_4, \ \ {\rm etc.}
\ear   
   Since $\kappa \approx l_{\rm Pl}^2$ (the Planck length squared), it is evident that our system 
   remains on the semiclassical scale only if all parameters \rf{ABC} are much smaller than unity. 
   Hence, in particular, the minimum spherical radius $r = r_0$, reached at bounce,
   should be much larger than the Planck length. Other parameters that should be small
   are values of the derivatives $\ddot \beta, \ddot\gamma$, etc. close to the bounce. 
      
   An inspection shows that, in the approximation used, it is sufficient to consider the order $O(1)$
   in the ${0 \choose 0}$ component of \eqs \rf{EE}, from which we find
\bearr                                                                   \label{00-0}
   	A = N_1 [-2 A^2+2(2B_2+C_2)^2] + N_2 [-A^2 +(B_2+C_2)^2+2B_2^2].
\ear
  The role of all other equations reduces to expressing the constants $B_4, C_4$, etc. in terms of 
  $A, B_2, C_2$.  Thus we have a single equation for the three parameters $A, B_2, C_2$ of 
  the bouncing geometry, along with the coefficients $N_1, N_2$. Therefore, we have a broad
  space of possible solutions.
  
  As stated above, we must assume that $r_)$ is much larger than the Planck length 
  $ l_{\rm Pl} \sim \sqrt{\kappa}$, from which it follows that $A \ll 1$, or $A = O(\eps)$, 
  $\eps$ being a small parameter. We can also make the natural assumptions
  $B_2 = O(\eps)$ and $C_2 = O(\eps)$, which means that $\ddot\beta$ and $\ddot\gamma$ 
  are of the same order of magnitude as $1/r_0^2$. Then, since the r.h.s. of \eqn{00-0} is 
  $O(\eps^2)$ while the l.h.s. is $O(\eps)$, to provide the equality, we must require that $N_1$ 
  and/or $N_2$ should be large, of the order $O(1/\eps)$. 
  
  The remaining Einstein equations ${1 \choose 1}$ and ${2 \choose 2}$ at $\tau =0$ 
  then show that $B_4$ and $C_4$ are of the order $O(\eps^2)$ (see \rf{HH}), 
  therefore, the 4th order derivatives of $\beta$ and $\gamma$ are of a correct order of 
  smallness with respect to the Planck scale (see \rf{ABC}). Similar estimates are obtained
  for $B_6, C_6$, etc. if we analyze equations in the order $O(\tau^2)$, and so on.
  It can also be verified that the curvature invariants $R$, $R\mn R\MN$ and 
  $\cK\equiv R_{\mu\nu\rho\sigma}R^{\mu\nu\rho\sigma}$ are small at bounce ($\tau=0$)
  as compared to the Planck scale: 
\bearr \label{inv}   
    	R=  \frac{2}{r_0^2} + 4\beta_2 + 2\gamma_2 = O\Big(\frac{\eps}{\kappa}\Big),
\nnn   
        R\mn R\MN =\frac{2}{r_0^4}+\frac{4\beta_2}{r_0^2}+6 \beta_2^2 
                        +4 \beta_2\gamma_2+2\gamma_2^2=O\Big(\frac{\eps^2}{\kappa^2}\Big),
\nnn   
	\cK=\frac{4}{r_0^4}+8\beta_2^2+4\gamma_2^2=O\Big(\frac{\eps^2}{\kappa^2}\Big).
\ear

   Consider a numerical example for illustration. Assuming $N_1 =0$, 
   $N_2= 10^{10}$,  and $A = 10^{-10}$, a minimum radius $a$ is of $10^5$ Planck lengths.
   Since, by construction  (see \rf{abc} and \rf{ABC}), $B_2 > 0$ and $C_2 <0$, 
   we can assume for convenience $B_2 + C_2 =0$. As a result, from \eqn{00-0} we find
\[
		B_2 = - C_2 = 10^{-10}.
\]            
  If we substitute this into the ${1 \choose 1}$ and ${2 \choose 2}$ components of the 
  Einstein equations at $\tau =0$, with the expressions \rf{Gmn} and \rf{Hmn}, we can obtain 
  the values of $B_4$ and $C_4$:
\[
  		B_4 = 3.5 \times 10^{-20}, \qquad C_4 = - 8.5 \times 10^{-20}.
\]
  From the equations of order $O(\tau^2)$ one can then determine $B_6, C_6$, and so on.

  One can recall that in \sph\ space-times, if the spherical radius $\e^\beta = r$ has a regular 
  minimum (it is a wormhole throat if the minimum is in an R-region and a bounce if it is in 
  a T-region), then the SET must satisfy the condition  $T^0_0 - T^1_1 < 0$ which means violation 
  of the Null Energy Condition, see, e.g.,  \cite{BR-book, BKor-15}. In our model, supposing
  a bounce at $\tau =0$, we automatically obtain the inequality $T^0_0 - T^1_1 < 0$.
  
\section{Nonlocal contribution to the vacuum SET}

   To estimate the contribution of the nonlocal term $P\mN$ in the SET \rf{SET}, we rewrite 
   the general metric \rf{ds0} of a \KS\ cosmology as
\beq
	ds^2 = \e^{2\alpha} d\eta^2 - \e^{2\gamma}dx^2 - \mu^2\e^{2\beta}d\Omega^2,
\eeq
   where the time coordinate $\eta$ is the so-called ``conformal time,'' defined by the condition 
   $3\alpha(\eta) = 2\beta(\eta) + \gamma(\eta)$, being convenient for considering quantum fields. 
   We assume that the \bh\ has a stellar (or larger) mass $m_{\rm Sch}$, and 
   $\mu=2Gm_{\rm Sch}\gtrsim 10^5$\,cm = 1 km is the corresponding gravitational radius. 
   Meanwhile, at bounce (say, at the time $\eta=0$), in agreement with the previous section, 
   we assume that the minimum radius is $r_0 = \mu \e^{\beta(0)}$ is  $\sim 10^5 l_{\rm Pl}\sim 10^{-28}$\,cm.
   Introducing the small parameter $\ep =r_0/\mu \lesssim 10^{-33}$ (not to be confused with $\eps$
   from the previous section), for times close to the bounce we can write
\beq  				\label{exps}
	\e^{2\alpha} = \ep(1+a \eta^2),  	\qquad
	\e^{2\beta} = \ep^2(1+b \eta^2),	\qquad
	\e^{2\gamma} = \ep^{-1}(1+c \eta^2).
\eeq
  with $3a = 2b + c$ according to the definition of $\eta$, $b > 0$ since $\e^\beta$ has a minimum, and $c < 0$ 
  since $\e^\gamma$ has a maximum at $\eta =0$. The powers of $\ep$ correspond to magnitudes 
  of the metric coefficients at approach to a would-be \Swz\ singularity.

  Consider a quantum scalar field satisfying the equation $(\Box + M^2 + \xi R) \Phi =0$ and its
  standard Fourier expansion:
\beq
	\Phi={\cal N} \e^{-\alpha}\int dk \sum_{lm}\e^{-ikx} Y_{lm}(\theta, \varphi)g_{klm}(\eta) c^{+}_{klm}+{\sf h.c.},
\eeq
  where ${\cal N} $ is a normalization factor, $\xi$ is a coupling constant, $ c^{+}_{klm}$ is a creation operator, 
  $Y_{lm}$ are spherical functions, and each mode function $g_{klm}(\eta)\equiv g$ obeys the equation 
  obtained by separation of variables in the original Klein--Gordon-type equation:
\beq
		\ddot{g}+\Omega^2 g=0,
\eeq
  where the dot stands for $d/d\eta$, and $\Omega$ is the effective frequency:
\beq 		\label{Omega}
		\Omega^2=k^2\e^{2(\alpha-\gamma)}+\frac{l(l+1)+2\xi}{\mu^2}
		\e^{2(\alpha-\beta)}+M^2\e^{2\alpha}+\frac{2\xi (\dot\beta-\dot\gamma)^2}{3}+
		(6\xi-1)(\ddot\alpha+\dot\alpha^2).
\eeq
   At bounce time $\eta =0$ we have, due to standard normalization, $|g| \sim \Omega^{-1/2}$ and 
\beq		\label{Omega0}
	      \Omega^2(0) = k^2\ep^2+\frac{l(l+1)+2\xi}{\mu^2\ep} + M^2\ep + (6\xi-1)a.
\eeq

  Now, for estimation purposes, we will make a natural assumption, justified by experience 
  \cite{birrell, gmm},
  that particle production takes place most intensively at energies close to the curvature scale $\sim r_0^{-1}$.  
  This energy is of the order of the frequency $\bar\Omega(\tau)$ in terms of the proper cosmic time $\tau$ 
  related to our conformal time by $d\tau=\e^\alpha d\eta$. Therefore our assumption means 
  $\bar\Omega \sim 1/r_0$. Since $\e^\alpha \sim \sqrt{\ep}$, one has $\tau\sim\sqrt{\ep}\eta$, and 
  from the relation $\Omega\eta = \bar\Omega\tau$ we obtain $\bar\Omega=\Omega/\sqrt{\ep}$, so that
\beq 		\label{Omegabar0}
		\bar\Omega^2(0)=k^2\ep+\frac{l(l+1)+2\xi}{\mu^2\ep^2} + M^2+\frac{(6\xi-1)a}{\ep}.
\eeq
  It is of interest, at which values do the parameters of the model appreciably contribute to 
  $\bar\Omega^2$ having the order $\sim r_0^{-2} = (\mu\ep)^{-2}$. They are:
\beq                           \label{orders}
	k \sim \frac{1}{\mu^{2}\ep^{3}}\sim10^{45}\,{\rm cm}^{-1}\sim 10^{12}\,m_{\rm Pl}; \qquad
	 l,\xi\sim 1; \qquad     
	 M\sim \frac 1 {r_0};  \qquad 
	 a = \ddot\alpha(0) \sim \frac {\ep}{ r_0^2}.
\eeq
  Apparently, momenta $k$ strongly exceeding the Planckian value look meaningless, and we can conclude that 
  at reasonable (sub-Planckian) values of $k$, their contributions to $\bar\Omega$ are negligibly small. 

  Note that the result $a\sim \ep/r_0^2$ can be obtained in another way using the relations
\[  
     e^{2\alpha}=\ep^{-1}(1+ \tau^2/r_0^2)=\ep^{-1}(1+a\eta^2),  \qquad    \tau\sim\sqrt{\ep}\eta. 
\]  
  A similar analysis leads to $b, c\sim \ep/r_0^2$.  Furthermore, at small $\eta$ we can assume
\beq 			\label{OmegaTaylor} \wide
	\Omega\approx B + C \eta^2, \quad\  \text{where}\ \  
	B=\Omega(0)\sim\sqrt{\ep}/r_0,  \qquad C/B\sim (a,b,c)\sim \ep /r_0^2.
\eeq

  The energy density of created particles may be estimated using the standard technique of Bogoliubov
  coefficients. For the case of bounce-type metrics, the crucial Bogoliubov coefficient $\beta_{kl}$ can be 
  computed with necessary accuracy by using the formulas \cite{Branden} 
\beq                  \label{II}
	\beta_{kl}=\sqrt{\frac{I^{-}}{I^{+}}}\sinh\sqrt{I^{-}I^{+}},
	\quad I^{\pm}\equiv\int_{\eta_1}^{\eta}g^{\pm}(\bar\eta)d\bar\eta,
	\quad g^{\pm}\equiv\frac{\dot\Omega}{2\Omega}\exp\left(\pm 2i\int_{\eta_1}^\eta 
		\Omega(\bar{\eta})d\bar{\eta}\right),
\eeq
  where $\eta_1$ is the initial time at which, by assumption, $\beta_{kl}=0$ (that is, assuming a vacuum 
  state of the field, with no particles). Using \eq\eqref{OmegaTaylor} and making the assumption
  $B\eta\lesssim O(1)$ (which means that $\eta$ is not very far both from zero and from $\eta_1)$,  we 
  obtain
\bearr
	\int_{\eta_1}^\eta \Omega(\bar{\eta})d\bar{\eta} \approx 
			B\bar\eta+\frac{1}{3}C\bar\eta^3\Bigr|^\eta_{\eta_1}\approx  B (\eta-\eta_1),
\\ \lal
	g^{\pm}(\eta) \approx \frac{C\eta}{B}\e^{\pm 2i B(\eta-\eta_1)}
			\sim \frac{\eps\eta}{r_0^2}\e^{\pm 2i B(\eta-\eta_1)}.
\ear
  Now we can calculate the integrals $I^\pm$ involved in \rf{II} at times close to bounce ($\eta=0$):
\bearr
	I^{\pm}(\eta)\Bigr|_{\eta\to0}\sim \frac{\eps}{r_0^2}\int_{\eta_1}^0\eta d\eta \e^{\pm 2i  
		B(\eta-\eta_1)}=\frac{\eps}{r_0^2}\e^{\mp2iB\eta_1}
			\left[\frac{\e^{\pm2iB\eta}}{4B^2}(1\mp2iB\eta) \right]^0_{\eta_1}
\nnn
		\inch =\frac{1}{4}\left[\e^{\mp2iB\eta_1}-1\pm2iB\eta_1\right]\approx-\frac{1}{2}B^2\eta_1^2.
\ear
  Then, assuming $B\eta_1\lesssim O(1)$, we arrive at
\beq
	\beta_{kl}\sim I^{-}\sim-\frac{1}{2}B^2\eta_1^2, \qquad  
	        |\beta^2_{kl}|\sim\frac{1}{4}B^4\eta_1^4\lesssim O(1).
\eeq 
  Thus the energy density of created particles is
\beq             \label{rho}
		\rho_{\rm nonloc}=\langle T^0_0\rangle \sim \frac{1}{8\pi}
	\int dk \sum_l(2l+1)\frac{e^{-4\alpha}}{\mu^2}\Omega |\beta_{kl}|^2
		\sim\frac{10^5\sqrt{\ep}}{r_0^4}\sim \frac{10^{-11}}{r_0^4},
\eeq
  where we have employed the following approximate orders of magnitude for each factor in \rf{rho}, in 
  agreement with \rf{orders}:
  (i) $\int dk\sim 2m_{\rm Pl} = {10^5}/{r_0}$ since we integrate from $-m_{\rm Pl}$ to $+m_{\rm Pl}$; 
  (ii) $\sum_l (2l+1) \sim 10^2$, involving a few low multipolarities (since large multipolarities would mean
  too large mode energies);  (iii)  $e^{-4\alpha}/\mu^2 \sim {1}/{r_0^2}$; (iv) $\Omega \sim \sqrt{\ep}/r_0$; 
  (v) $|\beta_{kl}|^2 \sim 1$ as a very rough upper bound.
  
  A comparison of the estimate \rf{rho} with that of the local energy density contribution from vacuum 
  polarization obtained in the previous section and \cite{we18}, $\rho_{\rm loc}\sim 10^{10}r_0^{-4}$,   
  leads to $\rho_{\rm nonloc}/\rho_{\rm loc}\sim 10^{-21}$, and this value is still smaller if we 
  consider \bh\ masses larger than that of the Sun. We conclude that the nonlocal contribution 
  to the vacuum energy density due to particle production is negligibly small in the regime of
  semiclassical bounce, and a more accurate calculation including more physical fields of different spins  
  can hardly change this estimate too strongly.

\section{Matter accretion into a \Swz\ \bh} 
\subsection{CMB accretion}

  Black holes in the real Universe are surrounded by various kinds of matter: interstellar or intergalactic 
  gas, dust and stellar matter if the \bh\ gravity destroys approaching stars. Depending on specific
  astrophysical circumstances, the ambient matter may form an accretion disk or experience spherical 
  or close to spherical accretion. The falling matter crosses the horizon and should ultimately approach 
  the \bh\ singularity, if the latter really exists. Or, if the theory predicts a bouncing region instead of a 
  singularity, it is natural to ask: will the gravity of the accreted matter strongly change the geometry of the 
  bouncing region? Can it happen that this falling matter will destroy the bounce (whatever be its origin) 
  and restore the singularity? 
  
  We will try to answer this question for a \Swz\ \bh\ with a semiclassical bounce described in \cite{we18}
  and in the previous sections. Thus we assume that the space-time metric is approximately \Swz,
\beq                        \label{Sch}
		ds^2 = \Big(1 -\frac{2m}{r}\Big) dt^2 - \Big(1 -\frac{2m}{r}\Big)^{-1} dr^2 - r^2 d\Omega^2,
\eeq    
  everywhere except for a region close to bounce, that is, $r \lesssim  n r_0$, where, say, $n \lesssim 10$,
  and $r_0$ is the minimum radius at bounce.
  
  In this subsection, we consider spherical accretion of the kind of matter that exists anywhere in the Universe, 
  the Cosmic Microwave Background (CMB). Thus our calculation can correspond to an isolated \Swz\ 
  \bh\ in intergalactic space, surrounded by the CMB only, and the accretion consists in capture of CMB
  photons. It is thus a minimum possible environment of any \bh.
  At each point of the \bh's ambient space, there is a flow of photons to be captured: these are
  photons whose path gets into the so-called photon sphere with the radius $r_{\rm ph} = 3m$. Such photons 
  may be considered as those forming a radiation flow with the SET  
\beq                      \label{T-flow}
		T\mN = \Phi(r, t) k_\mu k^\nu,  \cm    k_\mu k^\mu =0,
\eeq    
  where the null vector $k^\mu$ is, in a reasonable approximation, radially directed, so that
\beq                       \label{k-r}
		k^\mu = (\e^{-\gamma}, - \e^{\gamma}, 0,0), \qquad  
		k_\mu = (\e^{\gamma}, \e^{- \gamma}, 0,0),
\eeq      
  where $\e^\gamma = \sqrt{1-2m/r}$. Then the conservation law $\nabla_\nu T\mN =0$ in the metric \rf{Sch}
  gives for $\Phi = \rho_{\rm flow}$ (the flow energy density)
\beq                                 \label{Phi}                            
		\Phi(r,t) = \frac{\Phi_0}{r (r-2m)}, \qquad   \Phi_0 = \const.
\eeq          

  The constant $\Phi_0$ should be determined by the CMB energy density and the \bh\ mass, taking into 
  account bending of photon paths in the \bh's gravitational field. Fortunately, there is no necessity to
  carry out such a computation anew: we can use, for example, the result obtained by
  Bisnovatyi-Kogan and Tsupko \cite{bis}. They showed that if a source of radiation is located 
  at $r = 10^4 m$ in \Swz\ space-time, then the \bh\ will capture radiation emitted inside a cone with 
  an angular radius $\alpha \approx 0.0298\deg \approx 5.203\ten{-4}$. If the source radiates isotropically,
  then the fraction $\Delta(r)$ of the emitted radiation energy captured by the \bh\ will be equal to the part 
  of the complete solid angle of $4\pi$ contained in the spot of $\pi \alpha^2$, that is, 
\beq       \label{Del-BK}
                     \Delta(r) = \pi \alpha^2 /(4\pi) = \alpha^2/4 \approx 6.768 \ten{-8}  \ \ {\rm for}\ \ r = 10^4\,m. 
\eeq    

  At $r = 10^4 m$ or larger, the space-time may be regarded approximately flat, therefore, due to flux
  conservation, the fraction $\Delta$ should be proportional to $r^{-2}$; on the other hand, since the area
  of a sphere from which the flux is collected, is $\propto m^2$, it should be also $\Delta \propto m^2$.
  As a result, we can write, using \rf{Del-BK},
\beq  
  		\Delta(r) \approx \frac {\Delta_0 m^2}{r^2}, \qquad  \Delta_0 = \const \ \ 
  				\then \ \ \Delta_0 = \frac {\Delta(r) r^2}{m^2} \approx 6.678. 
\eeq  
  On the other hand, at such distances from the \bh, the CMB can be safely regarded homogeneous 
  and isotropic, and we can conclude that the accretion flow will have the energy density 
\beq                        \label{Phi_0}
		     T^0_0 \approx \frac{\Phi_0}{r^2} = \Delta(r) \rho_{\rm CMB} 	= \frac {\Delta_0 m^2}{r^2}
		     \ \ \then \ \  \Phi_0 = \Delta_0 m^2 \rho_{\rm CMB}, 
\eeq      
  where the CMB density $\rho_{\rm CMB}$ is nowadays 
\beq			\label{rho-CMB}
		\rho_{\rm CMB} \approx 0.4 \ten{-12}\ {\rm erg\, cm^{-3}} \approx 1.41 \ten{-128}\, l_{\rm Pl}^{-4}, 
\eeq    
  where $\rho_{\rm Pl} = l_{\rm Pl}^{-4}$ is the Planck density.
  
  Thus we know the SET \rf{T-flow} with \rf{Phi} and \rf{Phi_0} in the external region of the \bh, but 
  the quantity  \rf{Phi} diverges at the horizon $r=2m$. This looks natural since in our static reference 
  frame the radiation is infinitely blueshifted at the horizon, where this reference frame in no more valid. 
  However, our purpose is to find out how this radiation behaves deeply beyond the horizon. To extend
  the expression \rf{T-flow} to $r < 2m$, let us transform it to the Kruskal coordinates valid at all $r$.
  To do that, it is convenient to use at $r > 2m$ the so-called tortoise radial coordinate 
\beq                     \label{r_*}
		r_* = r + 2m \ln \Big( \frac{r}{2m} -1\Big) \ \ \ \then \ \ \
				ds^2 = \Big(1 - \frac r{2m}\Big)(dt^2 - dr_*^2) - r^2 d\Omega^2  
\eeq      
  (note that $r_* \to - \infty$ as $r \to 2m$). This coordinate belongs to the same static reference frame,
  hence the flow energy density is $T^0_0 = \Phi$. However, the null vector $k^\mu$ is now, 
  instead of \rf{k-r},
\beq                       \label{k-r*}
		k^\mu = (\e^{-\gamma}, - \e^{-\gamma}, 0,0), \qquad  
		k_\mu = (\e^{\gamma}, \e^{\gamma}, 0,0),
\eeq      
  where, as before, $\e^\gamma = \sqrt{1-2m/r}$, and the nonzero covariant SET components have the form
\beq               \label{T-r*}
		T_{00} = T_{01} = T_{10} = T_{11} = \Phi \e^{2\gamma} = \frac {\Phi_0}{r^2},
\eeq          
  convenient for the transformation. 
  
  The Kruskal coordinates $R,T$, in which the metric has the form
\beq                     \label{ds-kru}
		ds^2 = \frac{32 m^3}{r} \e^{-r/(2m)} (dT^2 - dR^2) - r^2 d\Omega^2,
\eeq    
  are related to $r_*, t$ by
\beq                  \label{s-kru} 
		t = 2m \ln \frac{R+T}{R-T},  \qquad  r_* = 2m \ln \frac {R^2 -T^2}{4}.
\eeq    
  Using this, we transform $T\mn$ to the Kruskal coordinates and find the nonzero components
\beq                  \label{T-kru}  
		T_{TT} = T_{TR} = T_{RT} = T_{RR} = \frac {16 \Phi_0 m^2}{r^2 (R+T)^2}. 
\eeq      
  In \rf{ds-kru} and \rf{T-kru} the horizon $r_* = - \infty \mapsto  R^2 = T^2$ is a regular surface, the static 
  region $r > 2m$ corresponds to $R^2 > T^2$, while at $r < 2m$ instead of the coordinates $r^*, t$ or 
  $r, t$ we can introduce their counterparts  $x$ (analog of $t$) and $\tau$ (analog of $r^*$) by putting,
  for $ T > R > 0$ (the upper quadrant in Kruskal's diagram),
\beq                   \label{kru-KS}
                R =  \e^{\tau /(4m)} \sinh \frac x {4m}, \qquad
                T =  \e^{\tau /(4m)} \cosh \frac x {4m},
\eeq    
  so that the metric acquires the Kantowski-Sachs form
\beq                                 \label{Sch-KS}
	        ds^2 = \Big(\frac{2m}{r}-1\Big)(d\tau^2 - dx^2) - r^2 d\Omega^2
	          =  \Big(\frac{2m}{r}-1\Big)^{-1} dr^2 - \Big(\frac{2m}{r}-1\Big) dx^2 - r^2 d\Omega^2,  
\eeq
  the two timelike coordinates $r$ and $\tau$ being related by
\beq
		\tau = r + 2m \ln \frac{2m-r}{2m}.
\eeq    
  The horizon corresponds to $r = 2m$ or $\tau \to -\infty$, while the singularity $r=0$ occurs at  $\tau =0$.
  
  Using \rf{kru-KS}, we transform the tensor \rf{T-kru} to the Kantowski-Sachs coordinates $\tau, x$, obtaining
\beq               \label{T-KS}
		T_{\tau\tau} = T_{\tau x} = T_{x \tau} = T_{xx} = \frac {\Phi_0}{r^2}, 
\eeq    
  from which it follows that the energy density is 
\beq                  \label{rho-KS}
                  T^\tau_\tau = \rho_{\rm flow} = \frac {\Phi_0}{r (2m -r)}.
\eeq    
   We see that in the Kantowski-Sachs reference frame, in which the \Swz\ metric looks very similar
   to its usual appearance in the static region, the expression for $\rho_{\rm flow}$ also looks very similar.
   It is the density in the same reference frame that was used for describing the bounce and can thus be 
   compared with the vacuum polarization density 
   $\rho_{\rm vac} \sim 10^{10} r_0^{-4} \sim 10^{-10}\rho_{\rm Pl}$ at bounce. 
   
   Assuming that the internal \Swz\ metric \rf{Sch-KS} is the true metric up to $r \gtrsim r_0 \ll 2m$,
   using \rf{Phi_0} and \rf{rho-CMB}, we obtain for such small radii 
\beq
			\rho_{\rm flow} \approx \frac {\Phi_0}{2mr} 
					\approx \frac {\Delta_0 m \rho_{\rm CMB}}{2r},  
\eeq      
 and, since $\Delta_0$ is of the order of unity, we conclude that the flow density at small radii 
 is larger than $\rho_{\rm CMB}$ approximately by a factor of $m/r$. For a \bh\ of stellar mass,
 $m \sim 10^5$ cm and $r \sim r_0 \sim 10^5 l_{\rm Pl}$ this factor is $\sim 10^{33}$, so that,
 with $\rho_{\rm vac} \sim 10^{-10} \rho_{\rm Pl}$ and recalling \rf{rho-CMB}, we obtain  
 $\rho_{\rm flow}/\rho_{\rm vac} \sim 10^{-85}$. 
 
 This ratio will certainly be larger for heavier \bhs\ and for earlier epochs when $\rho_{\rm CMB}$ 
 was larger by a factor of $(a_0/a)^4$, where $a$ is the cosmological scale factor and $a_0$ its
 present value. Assuming the existence of supermassive \bhs\ with $m \sim 10^9$ solar masses
 at scale factors $a \sim 10^{-3}\,a_0$ (that is, at $z\sim 1000$, close to the recombination time),
 the above ratio gains 21 orders of magnitude, resulting in $\rho_{\rm flow}/\rho_{\rm vac} \sim 10^{-64}$.  
 
 We conclude that CMB accretion cannot exert any influence on the model dynamics at small radii
 close to bounce or a would-be singularity inside a \Swz\ \bh. Very probably, accretion of ambient 
 matter can  be much more important, and our next task is to estimate its impact.
 
\subsection{Dust accretion} 
   
  Matter falling onto a \bh\ has in general the form of a hot gas, but close to the horizon this gas is
  nearly in a state of free fall \cite{frol}, therefore the approximation of dust freely radially moving to
  the horizon looks quite adequate, and it is reasonable to assume that the same regime well
  describes its further motion in the T-region. 
  
  Thus we consider the \Swz\ space-time with the metric \rf{Sch} or, in terms of the tortoise coordinate 
  $r_*$, \rf{r_*}. In this metric, we consider matter with the SET 
\beq
		T\mN = \rho u_\mu u^\nu,
\eeq    
  where the components of the 4-velocity vector $u^\mu$ for radial motion may be written,
  in terms of the radial coordinate $r_*$, in the form
\beq
		u^\mu = (\e^{-\gamma}\sqrt{1+v^2}, - \e^{-\gamma} v, 0, 0), \qquad
		u_\mu = (\e^{\gamma}\sqrt{1+v^2},  \e^{\gamma} v, 0, 0),
\eeq    
  where $v = \e^{-\gamma} d r_*/ds$ ($s$ is proper time along the world line), so that $u_\mu u^\nu =1$. 
  
  We assume a steady infalling flow, so that both $\rho$ and $u^\mu$ in the R-region ($r > 2m$) depend 
  on $r$ only. Then the conservation law $\nabla_\nu T\mN$ has two nontrivial components:
\bearr
		(\rho v \sqrt{1+ v^2})' = - \rho v \sqrt{1+ v^2}\, (2 \beta' + 2 \gamma').
\nnnv
		(\rho v^2)' + \rho v^2 (2 \beta' + 2 \gamma') + \rho \gamma' =0,		
\ear    
  where the prime denotes $d/dr$, $\e^\gamma = \sqrt{1-2m/r}$, $\e^\beta = r$. Solving these equations 
  to find $\rho$ and $v$ as functions of $r$, we obtain\footnote
  	{Note that the expressions for $\rho$ and $v^2$ in terms of $\beta$ and $\gamma$ are valid not
    	only in the \Swz\ metric but in any \ssph\ metric written as 
	\[    
  			  ds^2 = \e^{2\gamma(x)} (dt^2 - dx^2) - \e^{2\beta(x)} d\Omega^2.
	\]  }
\bearr        \label{rho1}
		\rho = \frac {K \e^{-2\beta} }{E \sqrt{E^2 - \e^{2\gamma}}} = \frac{K}{r^2 E \sqrt{E^2 - 1 + 2m/r}}, 
		\qquad  E, K = \const,
\yyy		  \label{v}
		v^2 = E^2 \e^{-2\gamma} -1 = \frac {E^2 r}{r - 2m} -1. 
\ear
   Recalling that dust particles move along geodesics, one can independently obtain $v^2$ from the 
   geodesic equations which lead precisely to the expression \rf{v}, and the constant $E$ has the 
   meaning of conserved energy in the course of geodesic motion.   
  
   Now, our task is to follow the motion of the dust flow to the T-region. To do that, we again use the 
   transformation \rf{s-kru}, now for $T\mn = \rho u_\mu u_\nu$, and the result in the $(R,T)$ coordinates is
\bearr                 \label{Td-kru}
         T_{TT} = \frac{16 m^2 \rho (ER - T\sqrt{E^2 - \e^{2\gamma}})^2}{(R^2 - T^2)^2)},
\nnn         		
         T_{RT} = \frac{16 m^2 \rho ((R^2+T^2)E\sqrt{E^2 - \e^{2\gamma}})
         		- RT(2E^2 - \e^{2\gamma})}{(R^2 - T^2)^2},
\nnn
	 T_{RR} = \frac{16 m^2 \rho (ET - R\sqrt{E^2 - \e^{2\gamma}})^2}{(R^2 - T^2)^2)}.         
\ear         
   One can verify that these expressions lead to the correct expression for the SET trace, $T^\mu_\mu = \rho$.
   The expressions \rf{Td-kru} are valid in both R- and T-regions, even though in the T-region ($r < 2m$)
   we have $\e^{2\gamma} < 0$, so this notation should be perceived as a symbolic one.
   
   The next step is to use the transformation \rf{kru-KS} to the metric \rf{Sch-KS}, which results in 
\bearr 			\label{Td-KS}
	T_{\tau\tau} = \rho (E^2 - \e^{2\gamma}) = \rho (E^2 - 1 + 2m/r),
\nnn
	T_{\tau x} = - \rho \frac{R^4 + T^4}{(T^2 - R^2)^2},
\nnn
	T_{xx} = \rho E^2.
\ear			   
   It is again easy to verify the correctness of these expressions by confirming that $T^\mu_\mu = \rho$,
   now in the metric \rf{Sch-KS} in terms of $\tau$ and $x$. 
  
   With \rf{Td-KS} we find the following expression for the energy density of the dust flow in the T-region:
\beq
		T^\tau_\tau = \frac {r}{2m-r} T_{\tau\tau} = \frac{K \sqrt {E^2 - 1 + 2m/r}}{E r (2m -r)}.
\eeq      
   Let us estimate this quantity at $r \ll 2m$, assuming $E=1$ (which corresponds to zero velocity
   of dust particles at infinity):
\beq                                         \label{rho_E}
		\rho_E = T^\tau_\tau = \frac K {\sqrt{2m} r^{3/2}}.
\eeq
   The constant $K$ can be found if we know the dust density at some $r$ in the R-region. To this end, 
   we can recall that, according to \cite{frol} (page 324), under typical conditions the falling matter density is 
   $\rho \simeq (6 \ten{-12}\ {\rm g/cm^3}) (2m/r)^{3/2}$. Thus, say, at $r = 10 m$ we obtain 
   $\rho \sim 10^{-12}\ \rm g/cm^3$ which approximately equals $2 \ten{-106}\,\rho_{\rm Pl}.$\footnote
      		{$1\ {\rm g/cm^3} \approx 2\ten{-94} \rho_{\rm Pl}$.} 
    We thus have
\beq
		\rho\Big|_{r = 10m} = \frac {K}{\sqrt{2m}(10m)^{3/2}} \simeq 2\ten{-106} \rho_{\rm Pl}
		\quad \then \quad     K \simeq m^2 \ten{-104}\, \rho_{\rm Pl}.
\eeq      		      		
   With this value of $K$, let us estimate the dust energy density $\rho_E$ at the radius 
   $r=r_0  = 10^5\,l_{\rm Pl}$, the supposed bounce radius. According to \rf{rho_E},
\beq
		\rho_E\Big|_{r = 10^5\,l_{\rm Pl}} \approx \frac K{\sqrt{2m} r^{3/2}} 
				= \frac {10^{-104}}{\sqrt 2}\,\Big(\frac mr\Big)^{3/2} \rho_{\rm Pl}.
\eeq              
   For the \bh\ mass $m \approx m_\odot$, we have $(m/r)^{3/2} \approx 10^{50}$, so that 
\beq                                     \label{est-dust}
		\rho_E\Big|_{r = 10^5\,l_{\rm Pl}} \approx 10^{-52} \rho_{\rm Pl} \approx 10^{-42} \rho_{\rm vac}
\eeq     
  if we assume $\rho_{\rm vac} \approx 10^{-10} \rho_{\rm Pl}$.
  We conclude that the influence of the accretion flow on the hypothetic semiclassical bounce is quite
  negligible. The situation does not change if we assume, say, the initial dust density 5 orders of 
  magnitude larger and a supermassive \bh\ of $10^9 m_\odot$: we thus gain about 18 orders of
  magnitude in \rf{est-dust}, and there still remains a difference of 24 orders.  

\section {Conclusion}

   We have constructed a simple model \cite{we18} describing a possible geometry that can exist
   deeply inside a sufficiently large \bh\ at its sufficiently early stage of evolution, when the
   Hawking radiation is negligible due to its extremely low temperature, and one could not yet 
   feel the influence of quantum entanglement phenomena. The model is semiclassical in nature 
   and is governed by vacuum polarization leading to the emergence of quadratic curvature 
   invariants in the effective action. We have assumed that the free constants appearing at
   these invariants have values of the same order as in some well-known models of the 
   inflationary universe, and showed that the corresponding terms in the effective Einstein 
   equations lead to solutions in which the \Swz\ singularity is replaced by a regular bounce,
   ultimately leading to a white hole. 
   
   Furthermore, we have argued that other quantum effects such as the Casimir effect,
   caused by the spherical topology of a subspace in the \KS\ cosmology inside the 
   \bh, and particle production from vacuum caused by a nonstationary nature of the metric, 
   make only negligible contributions to the total effective SET and therefore cannot destroy 
   the bouncing geometry. The same has been shown for possible classical phenomena that
   could interfere, namely, accretion of different kinds of matter and its further motion to
   the \bh\ interior. It can be said that, in a sense, our simple bouncing model is stable under 
   both quantum and classical perturbations.  
   
   It would be of substantial interest to study how this model will be modified if Hawking 
   radiation at its early stages is included into consideration. Another subject of future studies
   can be concerned with using similar assumptions for black holes with charge and spin, 
   where the nature of singularities is quite different and where Cauchy horizons
   take place. As is mentioned in \cite{ash20b}, according to the stability analysis of Kerr 
   and Reissner-Nordstr\"om space-times, their Cauchy horizons are unstable under
   small perturbations, from which it follows that a generic black hole singularity must
   be null rather than spacelike as in the \Swz\ metric, and the analysis of such singularities 
   and their possible avoidance should be a promising field of research. 

\subsection*{Acknowledgments}

  This publication was supported by the RUDN University program 5-100. 
  The work is also supported by RFBR Grant No. 19-02-00346. The work of KB was also partly 
  performed within the framework of the Center FRPP supported by MEPhI Academic Excellence 
  Project (contract No. 02.a03. 21.0005, 27.08.2013).

\end{document}